\documentclass[prl,twocolumn,showpacs,preprintnumbers,amsmath,amssymb]{revtex4}
\usepackage{graphicx}
\usepackage{dcolumn}
\usepackage{bm}

\begin{document}

\title{Non-extensivity Parameter of Thermodynamical Model of Hadronic Interactions at LHC energies.}

\author{Tadeusz Wibig}
\affiliation{University of \L \'{o}d\'{z};\\ The Andrzej So\l tan Institute for Nuclear Studies, Uniwersytecka 5,
90-950 \L \'{o}d\'{z}, Poland.}
\email{wibig@zpk.u.lodz.pl}


\begin{abstract}
The LHC measurements above SPS and Tevatron energies 
give the opportunity to test predictions of non-extensive thermodynamical picture
of hadronic interaction to examine measured transverse momenta
distributions for new interaction energy range. 
We determined Tsallis model non-extensivity parameter for the hadronization process before
short-lived particles decayed and distort the initial $p_\bot$ distribution.
We have shown that it follows exactly smooth rise determined at lower energies below present LHC record.
The shape of the $q$ parameter energy dependence is consistent with expectations
and the evidence of the asymptotic limit may be seen.
\end{abstract}

\maketitle

The thermodynamical concept of hadronization process following the fifty years old Hagedorn ideas was 
recently successfully used in 
\cite{1} to describe multiplicities  
in $p \bar p$, $e^+e^-$, and $NN$ high energy interactions. However, the main problem of such treatment:
non-exponential tails of transverse momentum distributions remains. One way of 
their explanation is proposed in \cite{2} introducing the pre-hadronization 
dynamics of fireballs. Different, more general, solution is to extend the standard, 
Boltzmann statistics
following the Tsallis idea \cite{3}. The generalised, non-extensive statistics is recently often used
to describe different kinds of phenomena: from interstellar dynamics to cosmic ray 
source puzzle \cite{inne_tsallis}. 
Among them the high energy interaction phenomenology is one of the fields where the new 
statistics is very successful.

It was shown that such non-extensive statistics
works well for $p_\bot$ spectra in $e^+e^-$ \cite{4}, and could be used for $p\bar p$ and $pp$
interactions too \cite{5,jhep}. 
Here we would like to show that the results 
of the Tsallis generalisation of Becattini canonical thermodynamics works well when applied also to the LHC energies 2.36 and 7 TeV. 

In Refs. \cite{cms2} and \cite{cms7} the function of the form 
inspired by the Tsallis formula 
\begin{equation} 
\label{fitq27}
E\: {{{d^3 N_{\rm ch}} \over dp^3 }}~
\sim~ 
{{dN_{\rm ch}} \over {d\eta }}\: 
\left( \:1~+~{{(q-1) \: E_\bot } \over T} \: \right)
^{- {1 \over {q-1}}}
\end{equation}
was used to fit the $p_\bot$ data. The shape of the transverse momentum (invariant) distribution is well described
by the above formula with the values of the $q$ parameter of 1.13 and 1.15 (with the systematic 
uncertainty of order of 0.005) and $T$: 130 and 145 MeV for energies 2.36 and 7 TeV, respectively.

We would like to use the same transverse momentum data, but not only to adjust the simple and 
convenient formula parameters. We would like to try to say something about the mechanism of the hadronization
process. The fits of the form of Eq.\ref{fitq27} shows the direction 
one can follow, but the physical
conclusions should be based on a bit more detailed analysis of the thermodynamical picture.

The formalism we used in the present work is described in details in \cite{jhep}. 
The generalisation of the partition function of the canonical picture is in the replacement
of the exponential 
Boltzmann factor by the respective Tsallis formula:
\begin{equation}
\exp\left(- \:{E \over T}\right) ~\longrightarrow 
~\left(1+{\left(q-1\right) \over T} \: \sqrt{p^2+m_j^2}\right)^
{-\frac{1}{q-1}},
\label{x}
\end{equation}
where $q$ is called the non-extensivity parameter. The
transverse momentum distribution is given by the partition function $Z$
\begin{eqnarray}
f\left( p_T^2 \right){\rm d}p_T^2=\sum_j \sum \limits_{n=1}^\infty
(-1)^{n+1} \gamma^{s_j\;n}\:
(2J_j+1) {V \over 2 \pi^2 }\times \nonumber \\
\times \int {\rm d} p_L \ x_{j} 
{Z(Q^0 - nq_j) \over Z(Q^0)} \; {\rm d}p_T^2\ ~,
\label{pt2}
\end{eqnarray}
\noindent 
where $x_j$ is the statistical weight of the particular state (with classical Boltzmann or modified Tsallis factor) 
$\gamma^{s_j}$ is the strangeness suppression factor,
$J_j$ is the spin of the $j$-th type hadron (of quantum numbers $q_j$). 
$V$ denotes the hadronization volume.

\begin{figure}[ht]
 \begin{center}
    \includegraphics[height=8cm]{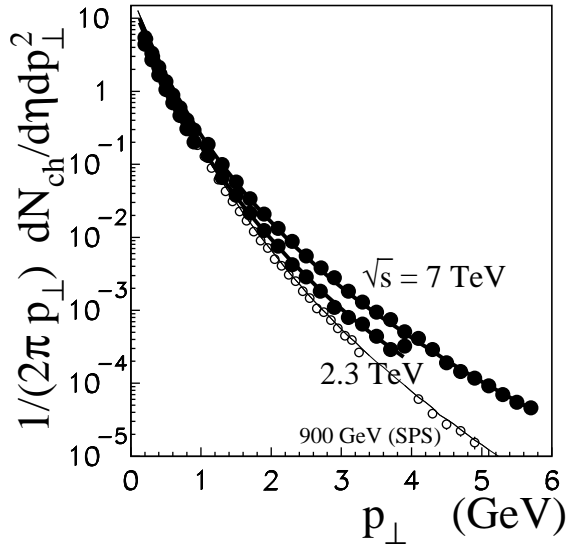}
  \end{center}
  \vspace{-0.5pc}
  \caption{ \label{pt}Transverse momentum distributions measured by CMS at 2.36 and 7 TeV and SPS at 
900 GeV and respective non-extensive statistics fits (described in the text).}
\end{figure}

Respective numerical integrations for 
the partition function determination 
have been performed. Then summation and integration of Eq.(\ref{pt2}) have been performed 
taking into account almost hundred
species of daughter particles. All unstable particles were then
decayed at the end and the final $p_\bot$ distribution was compared with the  distributions measured by CMS at two energies: 2.36 \cite{cms2}
and 7 TeV \cite{cms7}. The $T$ parameter is sensitive to small $p_\bot$'s
while the $q$ describes the non-exponential tails. 
No significant and systematic energy dependence of
$T$ were found, and eventually we found that this parameter can be fixed at 
130 MeV as in previous paper \cite{jhep}.
Results of the fits are shown in Fig.~\ref{pt}. The old data at 900 GeV form SPS 
experiment \cite{900} are shown together with the fit from \cite{jhep}.
 
\begin{figure}[ht]
  \begin{center}
    \includegraphics[height=9cm]{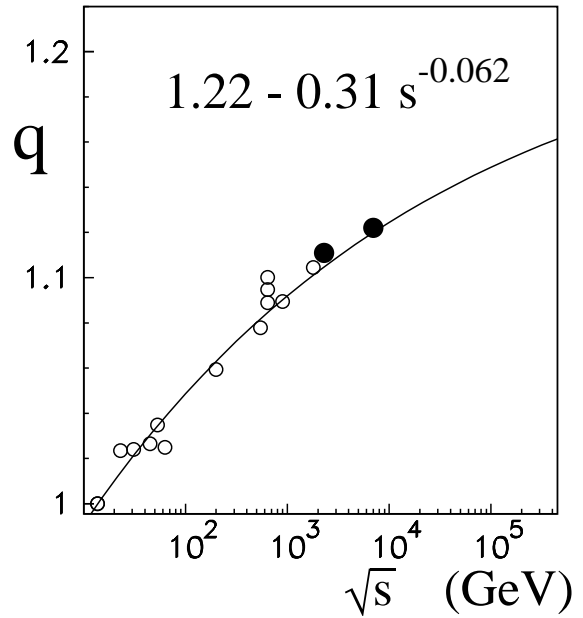}
  \end{center}
  \vspace{-0.5pc}
  \caption{\label{qfit}Energy dependence of the non-extensivity parameter. The open symbol represents 
the values obtained in \cite {jhep} for energies up to Tevatron. 
Two solid circles show values adjusted to CMS data.
}
\end{figure}

The non-extensive thermodynamics works well at lower energies, as it was shown e.g. in \cite{jhep} 
not only describing each individual energy transverse momentum spectrum, but the resulting 
parameter $q$ behaviour exhibits regular and expected smooth energy dependence. 
The low energy Boltzmann limit of $q=1$ (observed for 100 GeV/c data) together with 
the upper (asymptotic, for infinite energy) limit of 11/9 \cite{3} were combined in the functional form
\begin{equation}
\label{eqab}
q(E)=\left[\: 1.22~-~A\:s^{-B} \:\right]_{>1}   .
\end{equation}
\noindent
Parameters $A$ and $B$ were adjusted to the data up to Tevatron energies in \cite {jhep} and
the result is shown in Fig.\ref{qfit}.
The values of $A$ and $B$ are 0.308 and 0.062, respectively.

In the Fig.~\ref{qfit} new points at CMS energies are shown as solid circles. 

It is clearly seen that the 
non-extensive thermodynamical picture of hadron production applied to new LHC 
data on
transverse momentum distributions measured at 2.3 and 7 TeV c.m.s. energies gives 
good description of the experimental spectra.
The $T$ parameter could remain equal to 130 MeV and a smooth energy dependence of 
non-extensivity parameter $q$ found for lower energies is 
confirmed by the new data.

\end{document}